\documentclass[english,aps,reprint]{revtex4-1}
\usepackage[T1]{fontenc}
\usepackage[latin9]{inputenc}
\usepackage{amsmath}
\usepackage{amssymb}
\usepackage{graphicx}
\usepackage{esint}
\usepackage{babel}
\begin{document}
\title{The Bulk-Hinge Correspondence and Three-Dimensional Quantum Anomalous
Hall Effect in Second Order Topological Insulators}
\author{Bo Fu}
\affiliation{Department of Physics, The University of Hong Kong, Pokfulam Road,
Hong Kong, China}
\author{Zi-Ang Hu}
\affiliation{Department of Physics, The University of Hong Kong, Pokfulam Road,
Hong Kong, China}
\author{Shun-Qing Shen}
\email{sshen@hku.hk}

\affiliation{Department of Physics, The University of Hong Kong, Pokfulam Road,
Hong Kong, China}
\begin{abstract}
The chiral hinge modes are the key feature of a second order topological
insulator in three dimensions. Here we propose a quadrupole index
in combination of a slab Chern number in the bulk to characterize
the flowing pattern of chiral hinge modes along the hinges at the
intersection of the surfaces of a sample. We further utilize the topological
field theory to demonstrate the correspondent connection of the chiral
hinge modes to the quadrupole index and the slab Chern number, and
present a picture of three-dimensional quantum anomalous Hall effect
as a consequence of chiral hinge modes. The two bulk topological invariants
can be measured in electric transport and magneto-optical experiments.
In this way we establish the bulk-hinge correspondence in a three-dimensional
second order topological insulator.
\end{abstract}
\maketitle

\paragraph*{Introduction}

\begin{figure}
\includegraphics[width=8cm]{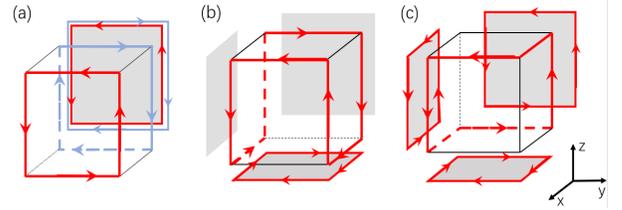}\caption{Illustration of selected patterns of chiral hinge modes and their
projection in a second order topological insulator in three dimensions.
(a) A double-loop pattern with the quadrupole indices $\Delta_{xy}=-\Delta_{zx}=1$
and $\Delta_{yz}=0$ and the slab Chern number $n_{x}=n_{y}=n_{z}=0$.
(b) A single-loop pattern with $\Delta_{xy}=1$ and $\Delta_{yz}=\Delta_{zx}=0$
and $n_{x}=n_{y}=0$ and $n_{z}=-1$. (c) A single-loop pattern with
$\Delta_{xy}=\Delta_{yz}=\Delta_{zx}=0$ and $n_{x}=n_{y}=-n_{z}=1$.\label{fig:Illustrationofhingemode}}
\end{figure}

The bulk-boundary correspondence lies at the heart of topological
states of matter and topological materials \citep{Prange-QHE,hasan2010colloquium,qi2011titsc,shen2012topological}.
It bridges the topology of bulk band structures and the physical observables
near the boundary. In the quantum Hall effect and quantum anomalous
Hall effect (QAHE), the quantized Hall conductance is associated with
the TKNN number of the band structure and the number of the edge modes
of electrons around the boundary \citep{TKNN-82prl,Laughlin-81prb,Wen-92IJMPB,hatsugai1993chern}.
In a topological insulator, a $Z_{2}$ index in the bulk is associated
with the number of the gapless Dirac cones of the surface electrons
\citep{kane2005z2,fu2007topoinv,ChenX16prb}. This reflects intrinsic
attributes of the topological phenomena. A recent advance in the field
of topological materials is the discovery of higher-order topological
insulators \citep{benalcazar2017quantized,schindler2018higher,schindler2018higherbi,benalcazar2017electric,song2017d,langbehn2017reflection,Khalaf2018higher,cualuguaru2019higher,ezawa2018higher}.
A second-order topological insulator in three dimensions refers to
an insulator with one-dimensional the chiral hinge modes (CHMs) localized
on the hinges at the intersection of adjacent side surfaces \citep{langbehn2017reflection,benalcazar2017electric,Khalaf2018higher,song2017d,geier2018second,schindler2018higher,schindler2018higherbi,fang2019new,trifunovic2019higher,yue2019symmetry,kooi2018inversion,van2018higher}.
Over the past few years, a great of efforts have been made to explore
the possible relation of the bulk bands and existence of hinge modes
as an extension of the bulk-boundary correspondence, such as effective
mass analysis \citep{song2017d,langbehn2017reflection,fang2019new,geier2018second,Khalaf2018higher,trifunovic2019higher},
the symmetry indicator \citep{kruthoff2017topological,po2017symmetry,khalaf2018symmetry,ono2018unified,watanabe2018structure,tang2019comprehensive,tang2019efficient,tanaka2020appearance,tanaka2020theory},
and spectral flow analysis \citep{takahashi2020bulk}. All the approaches
have their own merits. However, CHMs in a second order topological
insulator may display various flowing patterns as illustrated in Fig.
\ref{fig:Illustrationofhingemode}. It lacks a systematic method to
provide a comprehensive description of diverse flowing patterns. Also
it is desirable to learn which observable in the bulk is associated
with the CHMs.

In this Letter, we address the bulk-hinge correspondence and three-dimensional
(3D) QAHE as a physical consequence of the CHMs in a second-order
topological insulator. We start with a minimal four-band model to
reveal different flowing patterns of CHMs. It is found that a quadrupole
index is associated with the flowing direction of four hinge modes
of the system along one direction and a slab quantized Hall conductance
reveals the formation of a closed loop of the CHMs. We further demonstrate
the correspondent connection of the CHMs to the quadrupole index and
the slab Chern number by means of topological field theory. Finally
we propose to utilize magneto-optical Faraday and Kerr effects to
detect these topological invariants.

\paragraph{Model Hamiltonian and symmetry analysis}

We start with a minimal four-band Hamiltonian, $\mathcal{H}=\mathcal{H}_{0}+\sum_{i=1}^{3}\mathcal{V}_{i}$,
which consists of four parts. The first part is 
\begin{align}
\mathcal{H}_{0}= & \hbar\sigma_{x}[v_{\perp}(k_{x}s_{x}+k_{y}s_{y})+v_{z}k_{z}s_{z}]\nonumber \\
 & +[m_{0}+m_{\perp}(k_{x}^{2}+k_{y}^{2})+m_{z}k_{z}^{2}]\sigma_{z}s_{0}\label{eq:Hamiltonian}
\end{align}
where $k_{x},k_{y},k_{z}$ are the wave vectors, $m_{i}$ and $v_{i}$
are model parameters. $\boldsymbol{s}$ and $\boldsymbol{\sigma}$
are the Pauli matrices acting in spin and orbital space, respectively.
$\mathcal{H}_{0}$ possesses the time reversal symmetry $\mathcal{T}$
$(\mathcal{T}^{2}=-1)$ and belongs to the symplectic symmetry class
AII. Here we focus on the case of both $m_{0}m_{\perp}<0$ and $m_{0}m_{z}<0$
such that $\mathcal{H}_{0}$ describes a 3D strong topological insulator
with gapless Dirac cone of the surface states at all surfaces \citep{Shan10NJP,shen2012topological}.
$\mathcal{H}_{0}$ also respects the global chiral symmetry $\mathcal{C}=\sigma_{y}s_{0}$,
$\{\mathcal{C},\mathcal{H}_{0}\}=0$. Including the crystalline symmetries,
the total point symmetry group is $\mathcal{G}_{0}=D_{4h}\times\{1,\mathcal{T},\mathcal{P},\mathcal{C}\}$
with the particle-hole symmetry $\mathcal{P}\equiv\mathcal{C}\mathcal{T}^{-1}$\citep{Dresselhaus2008Grouptheory}.
As shown below all the terms in $\mathcal{H}$ preserve $\mathcal{P}$,
it is more convenient to rewrite $\mathcal{G}_{0}$ as $\mathcal{G}_{0}=\widetilde{\mathcal{G}}_{0}\times\{1,\mathcal{P}\}$
with the magnetic group $\widetilde{\mathcal{G}}_{0}=D_{4h}\times\{1,\mathcal{T}\}=D_{4h}\oplus\mathcal{T}D_{4h}$
(or $4/mm1^{\prime}$). $\mathcal{V}_{1}=c(k_{x}^{2}-k_{y}^{2})\sigma_{y}s_{0}$
breaks the time reversal symmetry $\mathcal{T}$. The presence of
$\mathcal{V}_{1}$ reduces magnetic group to $\widetilde{\mathcal{G}}^{\prime}=D_{2d}\oplus\mathcal{T}(D_{4h}-D_{2d})$.
The term proportional to $c$ opens an gap with opposite sign for
the surface states on the neighboring surfaces parallel to $z$ axis
and the CHMs may be localized at their intersections. The CHMs are
protected by the combination of four fold rotational symmetry and
time-reversal symmetry $\mathcal{R}_{4}^{z}\mathcal{T}$. The surface
states on the bottom $(00\bar{1})$ and top $(001)$ surface remain
gapless. $\mathcal{V}_{2}=d\sigma_{y}s_{0}$ and $\mathcal{V}_{3}=\sum_{i=x,y,z}b_{i}s_{i}$
is the magnetic Zeeman interaction. The two terms $d\sigma_{y}s_{0}$
and $b_{z}\sigma_{0}s_{z}$ anticommute with the linear terms $\hbar v_{\perp}(k_{x}\sigma_{x}s_{x}+k_{y}\sigma_{x}s_{y})$
along x and y directions. Thus they act as the mass terms and gap
out the surface states on $(001)$ and $(00\bar{1})$ while being
projected onto the x-y surface. Since both of them commute with the
mass term $\sigma_{y}s_{0}$, so they only modifies the mass term
for the surface states parallel with the z-axis and have no influences
on the four hinge states along the z direction. When all the surface
states are gapped out and the Fermi level is located in the surface
band gap, the electrons can only propagate unidirectionally along
the hinges shared by adjacent side surfaces due to time reversal symmetry
breaking. However, with different parameters, the chiral hinge modes
can exhibit distinctly different patterns. The presence of both $\mathcal{V}_{1}$
and $\mathcal{V}_{2}$ reduces the magnetic group to $\widetilde{\mathcal{G}}_{1}=D_{2}\oplus\mathcal{T}(D_{2h}-D_{2})$.
The term proportional to $d$ breaks both the time reversal $\mathcal{T}$
and inversion symmetry $\mathcal{I}$, respectively, but respects
the antiunitary combination $\mathcal{I}\mathcal{T}$ which means
the fact that if one CHMs propagates along any hinge there must be
another hinge state propagating in the same direction on its spatial
inversion. Thus the CHMs may form two closed loops on the surfaces
$(100)$ and $(\bar{1}00)$ as shown in Fig. 1(a). The relative sign
between $c$ in $\mathcal{V}_{1}$ and $d$ in $\mathcal{V}_{2}$
will determine which surface the two hinge mode loops locate around.

The presence of both $\mathcal{V}_{1}$ and $\mathcal{V}_{3}=b_{z}s_{z}$
with magnetic field in $z$ direction reduces the magnetic group to
$\widetilde{\mathcal{G}}_{2}=S_{4}\oplus\mathcal{T}(D_{2d}-S_{4})$.
The term breaks the $\mathcal{I}\mathcal{T}$ symmetry while preserving
the $\mathcal{S}_{4}$ symmetry which protects a single-loop CHMs
wriggling around the bulk as shown in Fig. 1(b). The relative sign
between $b$ and $c$ determines the wriggling way of the single-loop
CHMs. Only in the presence of $\mathcal{V}_{3}=b\sum_{i=x,y,z}s_{i}$
that magnetic field points to (111) direction, the magnetic point
group is $\widetilde{\mathcal{G}}_{3}=C_{i}\oplus\mathcal{T}(C_{2h}-C_{i})$.
Due to the presence of the inversion symmetry $\mathcal{I}$, the
CHMs at the inversion symmetric hinges are propagating in the opposite
directions, and form a closed loop as shown in Fig. 1(c). A detailed
symmetry analysis can be found in Ref. (\citep{Note-on-SM}).

\paragraph{Quadrupole index and slab Chern number}

In order to characterize the topological hinge modes, we introduce
two topological invariants: the quadrupole index and slab Chern number.
There are the CHMs along four hinges in the z direction in the case
of Figs. \ref{fig:Illustrationofhingemode}(a) and (b). The energy
dispersions of the four hinge modes connect the conduction and valence
bands, and cross at $k_{z}=0$ (see Fig.S1 in \citep{Note-on-SM}).
For a specific $k_{z}$, $\mathcal{H}(k_{z})$ can be viewed as a
2D system in the x-y plane and there are four corner states. The existence
of corner states can be characterized by the quadrupole moment \citep{kang2019many,wheeler2019many,li2020topological},
\[
q_{xy}(k_{z})=\frac{1}{2\pi}\mathrm{Im}\log\left[\mathrm{Det}[U_{k_{z}}^{\dagger}Q_{xy}U_{k_{z}}]\sqrt{\mathrm{Det}Q_{xy}^{\dagger}}\right]
\]
where the matrix $U_{k_{z}}$ is constructed by the occupied lowest
energy states, $Q_{xy}=e^{2\pi i\hat{r}_{x}\hat{r}_{y}/L_{x}L_{y}}$,
$\hat{r}_{\alpha}$ are the position operators, and $L_{\alpha}$
are the lengths of the system in the $\alpha$ direction. Any anti-symmetry
$\mathcal{O}_{a}$ leaves $xy$ plane invariant $\mathcal{O}_{a}\mathcal{H}(k_{z})\mathcal{O}_{a}^{-1}=-\mathcal{H}(-k_{z})$
will put a constraint on the quadrupole moment $q_{xy}(k_{z})$: $q_{xy}(k_{z})+q_{xy}(-k_{z})=0$
or 1. At two high symmetry points $\Lambda_{z}=0$ or $\pi$, the
symmetry is restored, $\mathcal{O}_{a}\mathcal{H}(\Lambda_{z})\mathcal{O}_{a}^{-1}=-\mathcal{H}(\Lambda_{z})$,
and $q_{xy}(\Lambda_{z})$ must be quantized to $0$ or $\frac{1}{2}$
(see Ref. \citep{Note-on-SM}). Non-zero quantized $q_{xy}(\Lambda_{z})$
indicates the system topologically nontrivial and the existence of
four zero-energy corner states in the reduced 2D subspace. For example,
if $q_{xy}(k_{z}=0)=1/2$, then $q_{xy}(\pm\pi)=0$ or 1. In this
case, there exist CHMs which compensate for the difference of the
corner charges. Thus we can introduce a quadrupole index, 
\begin{equation}
\Delta_{xy}=\int_{0}^{2\pi}dk_{z}\partial_{k_{z}}q_{xy}(k_{z})
\end{equation}
to characterize the existence and the flowing direction of four CHMs.
For the double-loop case in Fig. \ref{fig:Illustrationofhingemode}(a),
we have $\Delta_{xy}=-\Delta_{zx}=1$ and $\Delta_{yz}=0$, which
are protected by the combination of chiral symmetry and the mirror
symmetry $\mathcal{C}\mathcal{M}_{\alpha}$ and the combination of
chiral symmetry and the time reversal symmetry $\mathcal{C}\mathcal{T}$.
For the single-loop case in Fig. \ref{fig:Illustrationofhingemode}(b),
we have $\Delta_{xy}=1$ and $\Delta_{yz}=\Delta_{zx}=0$. The quadrupole
index along the $z$ direction is protected only by $\mathcal{C}\mathcal{T}$
and along the $x$ ($y$) is protected by both $\mathcal{C}\mathcal{M}_{x(y)}$
and $\mathcal{C}\mathcal{T}$. For the case in Fig. \ref{fig:Illustrationofhingemode}(c),
$\Delta_{xy}=\Delta_{yz}=\Delta_{zx}=0$.

The slab Chern number is another topological invariant as the quadrupole
index alone are not enough to characterize the diversity of the flowing
pattern of the CHMs. Consider a slab geometry of the sample with a
finite thickness $L_{z}$ with the periodic boundary condition along
the x and y direction. Denote the Bloch eigenstates by $|u_{n}(\boldsymbol{k}_{\perp},z)\rangle$
are the Bloch eigenstates, $\mathcal{H}(\boldsymbol{k}_{\perp},z)|u_{n}(\boldsymbol{k}_{\perp},z)\rangle=\varepsilon_{n}(\boldsymbol{k}_{\perp})|u_{n}(\boldsymbol{k}_{\perp},z)\rangle$
with $\boldsymbol{k}_{\perp}=(k_{x},k_{y})$ and the index $n$ for
the bands. The space-resolved Berry connection is given by $\mathcal{A}_{\alpha;n,n^{\prime}}(\boldsymbol{k}_{\perp},z)=-i\langle u_{n}(\boldsymbol{k}_{\perp},z)|\partial_{\alpha}|u_{n^{\prime}}(\boldsymbol{k}_{\perp},z)\rangle$
for the two occupied bands $n,n^{\prime}$. In this way we define
the slab Hall conductance and its relation to a slab Chern number
$n_{z}$\citep{tanaka2020theory}
\begin{equation}
\sigma_{xy}^{slab}=\intop_{0}^{L_{z}}dz\sigma_{xy}(z)=n_{z}\frac{e^{2}}{h}\label{eq:slabhall}
\end{equation}
where $\sigma_{xy}(z)=\frac{e^{2}}{2\pi h}\int d^{2}\boldsymbol{k}_{\perp}\mathrm{Tr}[\mathcal{F}_{xy}(\boldsymbol{k}_{\perp},z)]$
and $\mathcal{F}_{xy}(\boldsymbol{k}_{\perp},z)$ is the non-Abelian
Berry curvature in terms of $\mathcal{A}_{\alpha;n,n^{\prime}}(\boldsymbol{k}_{\perp},z)$.
Because of the periodicity of the Berry connection in the first Brillouin
zone, it can be proved that the slab Chern number $n_{z}$ is quantized
if the filled bands has a band gap to the excited states for a band
insulator. According to the bulk-boundary correspondence \citep{hatsugai1993chern},
each non-zero Chern number is associated with the closed loop of chiral
edge state. In Fig. \ref{fig:Illustrationofhingemode}(a), $n_{x}=n_{y}=n_{z}=0$,
while two quadrupole indices are not vanishing $\Delta_{xy}=-\Delta_{zx}=1$.
The system in a slab geometry (the open boundary condition is imposed
in the $y$ direction) is analogue to the quantum spin Hall insulator
except the the two counter-propagating hinges modes are localized
on the opposite sides. Experimentally, the quantized anomalous Hall
effect can be measured by using the surface-sensitive method \citep{chu2011surface}.
In Fig. \ref{fig:Illustrationofhingemode}(b), $n_{z}=-1$ and $n_{x}=n_{y}=0$.
There is a closed loop of chiral edge mode around the z axis. Combined
with the non-zero quadrupole index $\Delta_{xy}=1$. there are four
CHMs along the four hinges along the z axis, the two indices can determine
that a single-loop of CHMs that wriggles around the bulk. QAHE can
be detected through a global quantum Hall measurement probing the
whole sample due to the nonzero $n_{z}$. In Fig. \ref{fig:Illustrationofhingemode}(c),
$n_{x}=n_{y}=-n_{z}=1$. There is a single loop of chiral edge mode
around each axis. Because of the zero quadrupole indices around the
three axis, there is no four CHMs along one direction. It exhibits
a single CHM traversing half of its hinges, which can be projected
out a single closed loop in the direction of x, y and z. The QAHE
can be observed for three directions due to the non vanishing slab
Chern numbers.

\paragraph{3D QAHE}

The CHMs can be further understood in the framework of topological
field theory with an effective action\citep{qi2010chiral},
\begin{equation}
\mathcal{S}=\int d^{3}rdt\left[\frac{1}{8\pi}\left(\epsilon\mathbf{E}^{2}-\frac{1}{\mu}\mathbf{B}^{2}\right)+\frac{\theta(\mathbf{r},t)e^{2}}{4\pi^{2}\hbar c}\mathbf{E}\cdot\mathbf{B}\right],\label{eq:action}
\end{equation}
where $\mathbf{E}$ and $\mathbf{B}$ are the electromagnetic fields,
$\epsilon$ and $\mu$ are the dielectric constant and magnetic permeability,
respectively. $\theta(\mathbf{r},t)$ is known as the axion angle\citep{wilczek1987two}.
The product $\mathbf{E}\cdot\mathbf{B}$ is odd under the time reversal
or spatial inversion, $\theta$ has to be $0$ (modulo $2\pi$) for
a trivial insulator and the vacuum and $\pi$ for a topological insulator
with respect to the symmetries\citep{Note-on-SM,hughes2011inversion,turner2012quantized}.
In the quadratic order of electric and magnetic fields, besides the
Maxwell term, the $\theta$ term may give rise to the topologically
magneto-electric effect that an electric field can induce a magnetic
field and vice verse\citep{qi2010chiral,maciejko2010topological,tse2010giant,okada2016terahertz,dziom2017observation}.
By taking the functional derivative of $\theta$ term with respect
to a gauge field, the induced electric current density depends on
the spatial and temporal gradients of the $\theta$-field\citep{wilczek1987two,qi2010chiral},
\begin{equation}
\mathbf{j}_{\theta}(\mathbf{r},t)=\frac{e^{2}}{2\pi h}\left[\partial_{t}\theta(\mathbf{r},t)\mathbf{B}-\nabla\theta(\mathbf{r},t)\times\mathbf{E}\right].\label{eq:topological_current}
\end{equation}
The first term depends on the temporal gradient of the $\theta$-field
and is proportional to magnetic field, i.e., the so-called chiral
magnetic field, and vanishes in a static limit. The second term depends
spatial gradient of the $\theta$-field and is perpendicular to the
electric field, i.e., the anomalous Hall effect. Thus there will be
surface anomalous Hall effect at the interface between two regions
with different $\theta$ values and no Hall response will exist in
the bulk as $\theta$ takes a constant value $\theta_{b}$\citep{sitte2012topological,wang2015Quantized}.
The value of $\theta_{b}$ is given by the three-dimensional integration
of the Chern-Simons 3-form over momentum space\citep{qi2008topological,essin2009magnetoelectric}.
In addition to the inversion or time-reversal symmetry, $\theta_{b}$
will be quantized with improper rotation symmetries or a combination
of time-reversal symmetry and proper rotation symmetries\citep{fang2012bulk,Note-on-SM}.
From Eq. (\ref{eq:topological_current}), the layer-resolved Hall
conductivities in the $xy$ plane is associated with the gradient
of $\theta$, $\sigma_{xy}(z)=\frac{e^{2}}{2\pi h}\partial_{z}\theta(z)$.
Thus the slab Hall conductance (\ref{eq:slabhall}) is given by the
difference of the $\theta$ values of bottom and top vacuum $\sigma_{xy}^{slab}=\frac{e^{2}}{h}\frac{\theta_{T}-\theta_{B}}{2\pi}$,
which is integer-quantized independent of the $\theta$ value of the
bulk. 

\begin{figure}
\includegraphics[width=8cm]{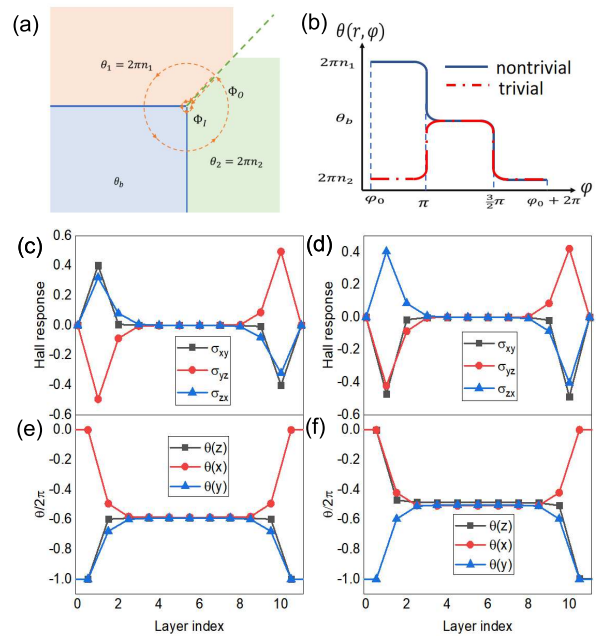}

\caption{(a) Schematic view of the hinge current. The planar surfaces of the
topological insulator are characterized by integers $n_{1}$ and $n_{2}$,
describing the integer change of the $\theta$ value nearby the surfaces.
(b) Schematic view of $\theta$-term as a function of the angle $\varphi$
for topologically nontrivial and trivial cases. (c) and (d) Plots
of the layer-resolved Hall response $\sigma_{\alpha\beta}(r_{\gamma})$
and (e) and (f) plots of the $\theta$-angle as function of the layer
index for two cases phases from a layer-resolved Kubo formula in a
slab geometry for $10$ layers.\label{fig:hall_theta}}
\end{figure}

\paragraph{Relation between the $\theta$ term and the chiral hinge modes}

The current carried by the CHMs can be evaluated from the spatial
dependent $\theta$, and each chiral hinge channel carries one conductance
quantum ($e^{2}/h$). We calculate the current through a 2D section
disk ($D$) encircling a hinge normal to the plane as illustrated
in Fig. \ref{fig:hall_theta}(a), $I=\iint_{D}d\boldsymbol{S}\cdot\boldsymbol{j}_{\theta}$.
The electric field is determined by the gradient of a scalar potential,
$\boldsymbol{E}=-\nabla\Phi(\boldsymbol{r})$, and we choose the boundary
of the disk as an equipotential line $\Phi_{e}$. By utilizing Stokes
theorem, $I=\frac{e^{2}}{2\pi h}\ointclockwise_{C}d\boldsymbol{s}\cdot\nabla\theta(\boldsymbol{r})\Phi(\boldsymbol{r})$.
Thus there is no current or equivalently gapless conducting channel
on the hinge when $\theta(\mathbf{r})$ in the two vacuum areas takes
the same value $n_{1}=n_{2}$. If they are different $n_{1}\ne n_{2}$,
there will be a branch cut separating the two vacuums where $\theta(\boldsymbol{r})$
is singular. In this situation, the contour integral gives the number
of the conducting channels $I/(\Delta\Phi e^{2}/h)=n_{2}-n_{1}$ which
is the winding number of the field $\theta(\boldsymbol{r})$. $\Delta\Phi=\Phi_{e}-\Phi_{in}$
denotes the potential difference between the outer contour $C_{out}$
and the inner contour $C_{int}$. In other words, the gapless hinge
mode tracks the singularity of the $\theta$ term and vice versa.
We also want to emphasize that, even when $\theta$ in the bulk is
not quantized, the above argument for the gapless chiral hinge channel
is still valid.

As show in Fig. \ref{fig:hall_theta} (c-f), we plot the layer-resolved
Hall responses $\sigma_{\alpha\beta}(r_{\gamma})$ ($\epsilon_{\alpha\beta\gamma}=1$)
and the integrated value for $\theta(r_{\gamma})$ as a function of
the layer index for three directions. In numerical evaluation, we
consider a slab geometry with the periodic boundary in the $\alpha\beta$
plane and open boundary condition in $r_{\gamma}$ direction. The
layer resolved Hall response only distributes near the slab surfaces
where $\theta$ changes and quickly drops to zero as the position
moves into the bulk where $\theta$ takes constant value. For the
double-loop case in Fig. \ref{fig:hall_theta} (c) and (e), the magnetic
point group $\widetilde{\mathcal{G}}_{1}$ will put a constraint on
the Hall response that the layer-resolved Hall conductivity takes
the opposite values for the slab center. Thus the slab Chern numbers
vanish for three directions. Due to the presence of the mass term
$d$, the axion angle will deviate form the quantized value $\pi$,
for example, $\theta_{b}/2\pi\simeq-0.59$ in Fig. \ref{fig:hall_theta}
(e). It is also consistent with the symmetry analysis that there is
no such symmetry to guarantee the quantization $\theta_{b}$ in $\widetilde{\mathcal{G}}_{1}$.
As a consequence, the surface Hall conductance $\sigma_{xy}^{B}=\frac{e^{2}}{h}(\frac{\theta_{b}}{2\pi}-n_{z}^{B})$
for the bottom interface and $\sigma_{xy}^{T}=\frac{e^{2}}{h}(n_{z}^{T}-\frac{\theta_{b}}{2\pi})$
for the top interface are not half quantized in sharp contrast to
the axion insulators. However, the summation of the surface Hall conductance
of the adjacent surface must be quantized since $\sigma_{zx}^{i}+\sigma_{zy}^{j}=\frac{e^{2}}{h}(n_{y}^{i}-n_{x}^{j})$
with $i,j=T,B$, indicates whether the hinge mode at the intersection
of two surfaces exists or not. For the single-loop case, the symmetry
$\widetilde{\mathcal{G}}_{2}$ constrains that the layer-resolved
Hall conductivities for $z$ direction are symmetric about the slab
center, while for $x$ and $y$ directions are antisymmetric about
the slab center. The layer-resolved Hall conductivities in $xz$ plane
and $yz$ plane are also related to each other by the $\mathcal{S}_{4}$
symmetry. Furthermore, $\theta_{b}$ will be quantized due to the
presence of improper rotation symmetry $\mathcal{S}_{4}$ and a combination
of time-reversal and the diagonal mirror symmetry $\mathcal{T}\mathcal{M}^{x+y}$.
As a result, the surface Hall conductance are half-quantized for three
directions. In this way, we establish the relation between the the
CHMs and the two physical invariants, $\sigma_{\alpha\beta}^{slab}=\frac{e^{2}}{h}(n_{\gamma}^{T}-n_{\gamma}^{B})$
and $\Delta_{\alpha\beta}=\delta_{n_{\beta}^{T}n_{\beta}^{B}}\delta_{n_{\alpha}^{T}n_{\alpha}^{B}}(n_{\alpha}^{T}-n_{\beta}^{T})$
with $\epsilon_{\alpha\beta\gamma}=1$.

\begin{figure}
\includegraphics[width=8cm]{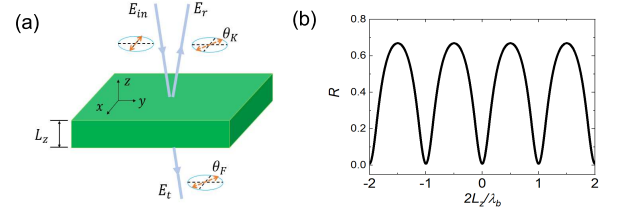}\caption{(a) Schematic illustration of the measurement of Kerr and Faraday
angle. Incident linearly polarized light becomes elliptically polarized
after transmission (Faraday effect) and reflection (Kerr effect),
with polarization angles as $\theta_{F}$ and $\theta_{K}$ respectively.
(b) The reflectivity $R$ as a function of the slab thickness $L_{z}$
along $z$ direction in the units of half of photon wavelength $\lambda_{b}/2$
for suspended single loop case with $\epsilon=10$ and $\mu=1$.}
\end{figure}

\paragraph{Magneto-optical effect as a detection of topological invariants}

Consider a normally incident linearly x-polarized light with frequency
$\omega$ propagating along the $z$ direction through the sample
$\mathbf{E}_{in}=E_{in}\exp[i(k_{0}z-\omega t)]\hat{\mathbf{x}}$
with $k_{0}=\omega/c$. $\mathbf{E}_{r}$ and $\mathbf{E}_{t}$ are
the reflected and transmitted electric field, respectively. Their
values at the interface between two materials are related to the incident
field $\mathbf{E}_{in}$ by the $2\times2$ reflection and transmission
tensors, and can be solved by matching the electrodynamic boundary
conditions. The Kerr and Faraday angles are defined by the $\tan\theta_{K}=-E_{r}^{y}/E_{r}^{x}$
and $\tan\theta_{F}=E_{t}^{y}/E_{t}^{x}$, respectively\citep{maciejko2010topological,Tse2011Magneto}.
When the chemical potential is located within the surface gap $E_{g}$
and $\hbar\omega\ll E_{g}$, the magnetic fields at the interface
of the two materials are discontinuous due to the presence of surface
Hall current. The reflection and transmission tensors for a slab can
be obtained by composing the single-interface scattering matrices
for top and bottom surfaces. For simplicity we only consider a free-standing
sample, the influence of a substrate do not change our conclusion
qualitatively. The reflectivity $R\equiv|\mathbf{E}_{r}|^{2}/|\mathbf{E}_{t}|^{2}$
will depend on the relative magnitude of the slab thickness and the
wavelength $(\lambda_{b}=\frac{2\pi c}{\omega\sqrt{\epsilon\mu}})$
inside the bulk. When the slab thickness contains an integer multiple
of half wavelength $L_{z}=N\lambda_{b}/2$ with an integer $N$ (the
resonance condition), $R$ reaches the minima. At the resonance, the
Faraday $\theta_{F}^{\prime}$ and Kerr $\theta_{K}^{\prime}$ rotations
have the same universal quantized value \citep{Note-on-SM,tse2010giant},
$\tan\theta_{F}^{\prime}=\cot\theta_{K}^{\prime}=\alpha(n_{z}^{T}-n_{z}^{B})$,
where $\alpha\equiv\frac{1}{4\pi\epsilon_{0}}\frac{e^{2}}{\hbar c}$
the fine structure constant. At the resonance, the difference of $\theta$
values between the top and bottom vacuum can be obtained irrespective
of the specific value of $\theta_{b}$. In order to determine $n_{z}^{T}-\frac{\theta_{b}}{2\pi}$
and $n_{z}^{B}-\frac{\theta_{b}}{2\pi}$ for top and bottom surface,
we also need to use the results at reflectivity maxima when $L_{z}=(N+\frac{1}{2})\lambda_{b}/2$.
The measured Faraday angle $\theta_{F}^{\prime\prime}$ and Kerr angle
$\theta_{K}^{\prime\prime}$ give a relation \citep{Note-on-SM}
\[
\tan(\theta_{K}^{\prime\prime}+\theta_{F}^{\prime\prime})\left(1-\frac{\tan\theta_{F}^{\prime}}{\tan\theta_{F}^{\prime\prime}}\right)=\alpha(n_{z}^{T}+n_{z}^{B}-\frac{\theta_{b}}{\pi}).
\]
Using the two relations, we can determine the values of the quadrupole
indices and the slab Chern numbers.

In short, the quadrupole index in combination with the slab Chern
number can determine the flowing pattern of the CHMs, which gives
rise to a 3D QAHE in a second order topological insulator.
\begin{acknowledgments}
BF and ZAH are contributed equally. This work was supported by the
Research Grants Council, University Grants Committee, Hong Kong under
Grant No. 17301220, and the National Key R\&D Program of China under
Grant No. 2019YFA0308603.
\end{acknowledgments}

\end{document}